# Impact of Local Descriptors Derived from Machine Learning Potentials in Graph Neural Networks for Molecular Property Prediction

Ryoichi Uchiyama[a], Yuya Nakajima[b,c], Yuta Tanaka[c], Junji Seino[a,b,*]

[a] Department of Chemistry and Biochemistry, School of Advanced Science and Engineering, Waseda University, Tokyo, 169-8555, Japan
[b] Waseda Research Institute for Science and Engineering, Waseda University, Tokyo, 169-8555, Japan
[c] AI Innovation Department, ENEOS Holdings, Inc., 8, Chidoricho, Naka-ku, Yokohama, Kanagawa, 231-0815, Japan

E-mail: j-seino@aoni.waseda.jp

Abstract

In this study, we present a framework aimed at enhancing molecular property prediction through the integration of local descriptors obtained from large-scale pretrained machine learning potentials into three-dimensional graph neural networks (3D GNNs). As an illustration, we developed an EGNN-PFP model by integrating descriptors derived from the preferred potential (PFP) features, acquired through Matlantis, into an equivariant graph neural network (EGNN), and evaluated its effectiveness. When tested on the QM9 dataset, comprising small organic molecules, the proposed model demonstrated superior accuracy compared to both the original EGNN models and the baseline models without PFP-derived descriptors for 11 out of the 12 molecular properties. Furthermore, when evaluated on the tmQM dataset, which encompasses transition metal complexes, notable enhancements in performance were observed across all five target properties, indicating the significance of the local atomic environment surrounding transition metals. In essence, the proposed methodology is adaptable to any 3D GNN architecture, and further enhancements in prediction accuracy are anticipated when integrated with continually evolving GNN architectures.

1. Introduction

The precise prediction of molecular properties is a fundamental technology for expediting the identification of potential candidate compounds within extensive chemical spaces in areas such as drug discovery, catalyst design, and the investigation of functional materials. While experiments and first-principles calculations are very dependable, their computational expenses and scalability constraints make them challenging to utilize for issues encompassing vast search spaces. Consequently, machine learning models that swiftly estimate quantum-chemistry-based properties from molecular structures have gained significant popularity in recent years as an efficient approach to substantially enhance exploration efficiency.[1,2]

Molecular properties are inherently dependent on the three-dimensional (3D) structure. The vector properties of all molecules, such as dipole moments, and electronic quantities, such as the HOMO–LUMO gap, cannot be uniquely determined from two-dimensional (2D) topological information alone, such as atomic connectivity and bonding patterns. Instead, the 3D geometric arrangements, including interatomic distances, bond angles, and molecular conformations, play a decisive role. Therefore, in predicting molecular properties, it is essential to develop models that directly handle 3D coordinates and learn in a way that satisfies geometric symmetries, such as rotational and translational invariance.

To meet the specified requirements, graph neural networks (GNNs) incorporating 3D geometric information as input have undergone rapid development. SchNet[3] has shown that local atomic environments can be efficiently represented by continuous-filter convolutions relying on interatomic distances, facilitating the effective learning of quantum chemical properties such as energies and forces. It has become evident that beyond distance-based two-body interactions, three-body interactions involving angular information are crucial for enhancing prediction accuracy. Models such as DimeNet[4] explicitly incorporated directional information, resulting in significant enhancements in the precision of molecular property predictions. Furthermore, to effectively handle vector and tensor quantities such as dipole moments and polarizabilities, 3D models based on rotationally equivariant representations have been developed. PaiNN[5] has demonstrated enhancements in both accuracy and data efficiency for predicting tensor properties by employing equivariant message passing, which concurrently updates scalar and vector features.

Subsequently, these rotation-equivariant designs have been expanded to encompass broader theoretical frameworks. Tensor field networks (TFN)[6] provide a structured approach to manage scalars, vectors, and higher-order tensors through equivariant layers grounded in spherical harmonics. Concurrently, SE(3)-Transformer[7] enhances representational capacity by incorporating equivariant attention mechanisms. Considering the trade-off between implementation efficiency and expressive power, E(n)-equivariant GNNs (EGNNs)[8] introduce a design that achieves equivariance without the need for higher-order tensor representations and has found extensive application across diverse tasks, such as molecular property prediction.

Efforts have also been made to explicitly model the hierarchical nature of the intramolecular interactions. GemNet[9] achieves high-accuracy molecular representations by explicitly incorporating bond lengths, bond angles, and dihedral angles, thereby accounting for the interactions of up to four-body terms. PaxNet,[10] inspired by the distinction between local and nonlocal interactions in molecular mechanics, employs a two-layer structure based on multiplex graphs to process information at different distance scales separately. Moreover, it has demonstrated effectiveness in RNA 3D structure prediction. SphereNet[11] efficiently processes directional information using spherical-harmonic-based representations, whereas ComENet[12] realizes comprehensive 3D molecular representations via complete message-passing mechanisms. More recently, models such as ViSNet[13] have been proposed to enhance the quality of geometric representations by strengthening the vector-scalar interactions within equivariant message passing. Thus, 3D molecular learning has continuously advanced by incorporating information such as distances, angles, equivariance, higher-order interactions, and attention mechanisms while preserving geometric symmetries.

Despite these advances, the majority of existing 3D GNNs mainly utilize atomic species and geometric structural data as inputs. They operate under the assumption that details regarding local electronic states are inherently captured within the network. While this assumption may lead to accurate predictions in specific scenarios, it is uncertain whether comprehensive information about local electronic environments can be accurately deduced solely from geometric data, especially as the diversity of elements or local electronic states increases. Therefore, a systematic investigation into the impact of explicitly incorporating local information on predicting global molecular properties is essential for formulating fundamental guidelines for designing 3D GNNs.

Numerical descriptors that encode local atomic environments have been extensively studied to represent local information. Atom-centered symmetry functions[14] represent local structures using distance- and angle-based symmetry functions and have become standard descriptors in the field of neural network potential (NNPs). The smooth overlap of atomic positions[15] provides rotationally invariant representations of local environments based on atomic density overlaps and is widely used to assess the similarity between environments. Additionally, the many-body tensor representation[16] offers a unified representation that encodes many-body interactions as distributions. Simultaneously, the atomic cluster expansion (ACE)[17] formulates local environments as systematically expandable bases to arbitrary order, enabling principled improvements in local representation accuracy. The Faber–Christensen–Huang–Lilienfeld descriptor[18] has facilitated high-accuracy regression of quantum chemical quantities by refining atomic environment representations. While these local descriptors can accurately encode the structural and chemical environments around atoms, they have mainly been utilized as inputs for kernel regression, fully connected regression models, and potential learning. Consequently, the systematic integration of such descriptors as node features in 3D GNNs for global molecular property predictions has not been thoroughly standardized yet.

In recent years, closely related to this line of research, there has been a growing interest in transferring internal representations learned by machine learning potentials pretrained on large-scale first-principles datasets to downstream property prediction tasks. ANI-1[19] achieved high-accuracy force fields by learning from large datasets, while DeepMD[20] enabled scalable molecular dynamics simulations. PhysNet[21] provides a framework for simultaneously predicting dipole moments and partial charges. Preferred potential (PFP)[22] features were proposed as a general NNP applicable to many elements, and the internal representations assigned to each atom during training can be interpreted as high-dimensional vectors encoding information regarding local environments and interactions. Similarly, equivariant potential models such as NequIP,[23] MACE,[24] and Allegro[25] have shown that information-rich local representations can be learned through tensor products and Clebsch–Gordan operations.

In this study, we focused on a framework that combines GNNs designed for 3D structures with pretrained local descriptors to predict global molecular properties. The main aim of this research is not to enhance a specific GNN architecture but to introduce a general design principle: providing pretrained local electronic representations as node features to GNNs capable of aggregating 3D geometric information. As an illustration, we utilized an EGNN as the GNN architecture and atomic representations obtained from the PFP as local descriptors; however, this integration can be easily extended to other 3D GNNs. The findings presented in this study elucidate how the combined effects of enriching local representations and aggregating geometric information, which are distinct from refining geometric representations, contribute to predicting global molecular properties.

2. Method

In this section, we describe the architecture and specifics of the EGNN-PFP model, which incorporates pretrained local descriptors from PFP into an EGNN to address 3D structures.

2.1 Overview of the EGNN-PFP Model

Figure 1 illustrates the overall architecture of the EGNN-PFP model. The model comprised three components: an input layer, an EGNN layer, and an output layer. Within the input layer, molecular graphs were built, and node and edge features were initialized. The EGNN layers updated atomic features and geometric information via equivariant message passing. The output layer executed task-specific molecular property predictions.

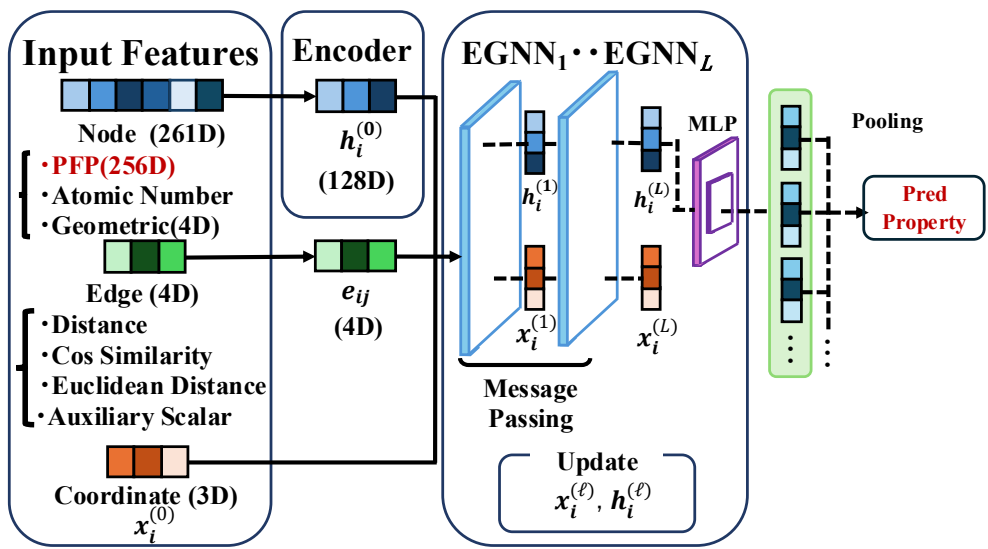

Figure 1: Architecture of the EGNN-PFP model.

2.2 Input Layer

In EGNN, every molecule is depicted as an undirected graph. This graph comprises a pair $G = (V, E)$, where $V$ is the set of vertices, and $E$ is the set of edges; the vertices correspond to the atoms while the edges signify the interactions between them. Typically, EGNN implementations utilize a fully connected graph linking all atom pairs, which escalates computational expenses notably for extensive molecules. Therefore, in this study, a fully connected graph was adopted for small molecules, while a graph was structured according to the cutoff distance for larger molecules.

When a cutoff is applied, an edge is constructed only for atomic pairs whose Euclidean distance $\|x_i - x_j\|$ is within the cutoff distance $r_{cut}$:

$$E = \{(i,j) \in V \times V \mid i \neq j, \|x_i - x_j\| \leq r_{cut}\}. \tag{1}$$

Here, $(i,j)$ denotes an edge connecting atoms $i$ and $j$.

Each atomic node $i \in V$ is represented by an initial feature vector $\boldsymbol{h}_i^{(0)} \in \mathbb{R}^{261}$. This vector is constructed by concatenating a local descriptor derived from PFP $\boldsymbol{p}_i^{\text{PFP}}$, atomic number $z_i$, and geometric features $\boldsymbol{g}_i$:

$$\boldsymbol{h}_i^{(0)} = \left[\boldsymbol{p}_i^{\text{PFP}} \oplus z_i \oplus \boldsymbol{g}_i\right]. \tag{2}$$

Here, $\oplus$ denotes the concatenation operator. The descriptor $\boldsymbol{p}_i^{\text{PFP}}$ is a 256-dimensional vector extracted from the final hidden layer of the NNP immediately before the energy prediction. The geometric feature vector $\boldsymbol{g}_i$ consists of the following four components: (i) the Euclidean distance $\|\boldsymbol{x}_i - \boldsymbol{x}_{\text{center}}\|$ between atom $i$ and the molecular center of mass $\boldsymbol{x}_{\text{center}}$; (ii) the number of neighboring atoms within the cutoff distance $r_{cut}$; (iii) the local atomic density obtained by normalizing the number of neighboring atoms by the cutoff sphere volume $4\pi r_{\text{cut}}^3/3$; and (iv) the coordinate norm $\|\boldsymbol{x}_i\|/d_{\max}$, normalized by the maximum interatomic distance $d_{\max}$ in the molecule.

Each edge $(i,j) \in E$ is represented by a feature vector $\mathbf{e}_{ij}$ that encodes spatial relationships and electronic similarity. In this study, a four-dimensional edge feature vector was utilized:

$$\boldsymbol{e}_{ij} = \left[d_{ij}, s_{ij}^{\cos}, \|\boldsymbol{p}_i^{\text{PFP}} - \boldsymbol{p}_j^{\text{PFP}}\|, I_{ij}\right]. \tag{3}$$

Here, $d_{ij}$ denotes the Euclidean distance between the atoms. The cosine similarity of the PFP descriptors, $s_{ij}^{\cos}$, is defined as:

$$s_{ij}^{\cos} = \frac{\boldsymbol{p}_i^{\text{PFP}} \cdot \boldsymbol{p}_j^{\text{PFP}}}{\|\boldsymbol{p}_i^{\text{PFP}}\|\|\boldsymbol{p}_j^{\text{PFP}}\| + \varepsilon}, \tag{4}$$

where $\varepsilon$ is a small constant introduced for numerical stability to avoid division by zero. The interaction term $I_{ij}$, which reflects both the spatial proximity and electronic similarity, is defined as

$$I_{ij} = s_{ij}^{\cos} \times d_{ij}. \tag{5}$$

With this definition, atom pairs that are spatially close and electronically similar are highlighted in the edge features.

2.3 EGNN Layers

In EGNN layers, atomic features are iteratively integrated with information from neighboring atoms through equivariant message passing. For the $\ell$-th layer ($\ell = 1, 2, \ldots, L$, with $L$ being the total number of layers), the processes for message passing and feature update include edge message computation, message aggregation, node update, and coordinate update.

First, the edge message $\boldsymbol{m}_{ij}^{(\ell)}$ is computed as

$$m_{ij}^{(\ell)} = \phi_e^{(\ell)}\left(\left[h_i^{(\ell)} \oplus h_j^{(\ell)} \oplus \|x_i - x_j\|^2 \oplus e_{ij}\right]\right), \tag{6}$$

where $\phi_e^{(\ell)}$ is a two-layer multilayer perceptron (MLP).

Next, for each node $i$, messages from the neighboring node set $\mathcal{N}(i)$ are aggregated. Here, $\mathcal{N}(i)$ denotes the set of all neighboring atoms connected to atom $i$ by the edges in $G = (V, E)$. Message aggregation is performed using either sum or mean aggregation.

$$m_i^{(\ell)} = \sum_{j \in \mathcal{N}(i)} m_{ij}^{(\ell)} \text{ or } \frac{1}{|\mathcal{N}(i)|} \sum_{j \in \mathcal{N}(i)} m_{ij}^{(\ell)}. \tag{7}$$

Node features are updated using residual connections:

$$h_i^{(\ell+1)} = h_i^{(\ell)} + \phi_h^{(\ell)}\left(\left[h_i^{(\ell)} \oplus m_i^{(\ell)} \oplus h_i^{(0)}\right]\right), \tag{8}$$

where $\phi_h^{(\ell)}$ is a two-layer MLP whose final layer does not include an activation function.

In the coordinate update step, the input molecular structure is refined to a geometry suitable for the task. When the coordinate updates are enabled, the atomic coordinates are updated as follows:

$$x_i^{(\ell+1)} = x_i^{(\ell)} + \sum_{j \in \mathcal{N}(i)} \frac{(x_i^{(\ell)} - x_j^{(\ell)})}{\|x_i^{(\ell)} - x_j^{(\ell)}\|} \phi_x^{(\ell)}\left(m_{ij}^{(\ell)}\right). \tag{9}$$

Here, $\phi_x^{(\ell)}$ is a two-layer MLP that takes the edge message $m_{ij}^{(\ell)}$ as input and outputs a scalar value representing the coordinate update based on the interaction between atoms. By omitting an activation function in the final layer, the coordinate update magnitude can take both positive and negative real values. When the coordinate updates are disabled, the coordinates remain unchanged as $x_i^{(\ell+1)} = x_i^{(\ell)}$.

Furthermore, edge messages are modified as attention mechanisms to evaluate edge importance dynamically.

$$m_{ij}^{(\ell)} \leftarrow m_{ij}^{(\ell)} \cdot \alpha_{ij}, \tag{10}$$

$$\alpha_{ij} = Sigmoid\left(\phi_{att}\left(m_{ij}^{(\ell)}\right)\right), \tag{11}$$

where $\phi_{att}$ is a two-layer MLP that takes the edge message $m_{ij}^{(\ell)}$ as input and outputs a scalar value representing the importance of the corresponding edge. Passing this value through a sigmoid function yields a weight coefficient $\alpha_{ij} \in (0, 1)$, which dynamically adjusts the contribution of each edge message. Through this attention mechanism, the model learns the relative importance of different interactions. For example, strong interactions, such as covalent bonds, may receive high weights. In contrast, in systems where weak nonbonded interactions (e.g., van der Waals forces) are important, appropriate weights can be assigned. This enables flexible representations to adapt to the chemical characteristics of each system.

2.4 Output Layer

The node features obtained after passing through the EGNN layers are utilized to predict the molecular-level properties in the output layer. This transformation comprised three stages: (i) node feature decoding, (ii) graph-level aggregation, and (iii) final prediction.

At the final EGNN layer (the $L$-th layer), the node feature $h_i^{(L)}$ of each atom is independently transformed by a node decoder $\phi_{\text{node}}$:

$$h_i^{\text{dec}} = \phi_{\text{node}}\left(h_i^{(L)}\right). \tag{12}$$

Here, $\phi_{\text{node}}$ is implemented as an MLP that maps the representations learned by the EGNN layers into a task-specific feature space. No activation function was applied to the final layer, resulting in a linear transformation.

In the graph-level aggregation step, the node features of all atoms are combined to create a singular vector representing the entire molecule. Depending on the nature of the target property and dataset characteristics, either sum or mean pooling was utilized.

$$h_{\text{graph}} = \sum_{i=1}^{N} h_i^{\text{dec}} \text{ or } \frac{1}{N}\sum_{i=1}^{N} h_i^{\text{dec}}. \tag{13}$$

Finally, the graph-level feature vector $h_{\text{graph}}$ is mapped to the final predicted property value $y$ through a graph decoder $\phi_{\text{graph}}$:

$$y = \phi_{\text{graph}}(h_{\text{graph}}). \tag{14}$$

Here, $\phi_{\text{graph}}$ is implemented as an MLP that applies nonlinear activation functions in the intermediate layers and a linear transformation without an activation function in the final layer, enabling prediction of arbitrary real-valued molecular properties.

3. Computational Details

3.1 Dataset and Model Configuration

To assess the precision of molecular property prediction utilizing the EGNN-PFP model, we utilized two quantum chemical datasets: QM9, comprising 133,831 organic molecules,[26] and tmQM, encompassing 86,000 transition-metal complexes.[27] The specific properties forecasted for each dataset are outlined in Table 1, while the primary distinctions in dataset structure are detailed in Table 2.

Table 1: Molecular properties predicted in this study.

| Dataset | Property | Description (Unit) |
|---|---|---|
| QM9 | $\mu$ | Dipole moment (Debye) |
| | $\alpha^3$ | Isotropic polarizability (Bohr³) |
| | $\varepsilon_{HOMO}$ | HOMO energy (meV) |
| | $\varepsilon_{LUMO}$ | LUMO energy (meV) |
| | $\Delta\varepsilon$ | HOMO-LUMO gap (meV) |
| | $<R^2>$ | Electronic spatial extent (Bohr$^2$) |
| | ZPVE | Zero-point vibrational energy (meV) |
| | $U_0$ | Internal energy at 0 K (meV) |
| | $U$ | Internal energy at 298.15 K (meV) |
| | $H$ | Enthalpy at 298.15 K (meV) |
| | $G$ | Free energy at 298.15 K (meV) |
| | $C_v$ | Heat capacity at 298.15 K (cal/mol·K) |
| tmQM | $\mu$ | Dipole moment (Debye) |
| | Metal_q | Partial charge on a metal atom |
| | $\varepsilon_{HOMO}$ | HOMO energy (meV) |
| | $\varepsilon_{LUMO}$ | LUMO energy (meV) |
| | $\Delta\varepsilon$ | HOMO-LUMO gap (meV) |

Table 2: Model design and hyperparameters for each dataset.

| Parameter | QM9 | tmQM |
|---|---|---|
| [Model Design] | | |
| PFP Calculation Mode | ωB97XD | r2SCAN |
| Edge Feature Dimension | 4 | 5 |
| Message Aggregation | Sum | Mean |
| Coordinate Update | OFF | ON |
| Graph Pooling | Sum | Sum($\mu$), Mean(others) |
| [EGNN Architecture] | | |
| Hidden Dimension | 128 | 128 |
| EGNN Layers | 7 | 7 |
| Node Decoder Layers | 2 | 2 |
| Node Decoder Dimensions | 128→128→128 | 128→128→128 |
| Graph Decoder Layers | 2 | 2 |
| Graph Decoder Dimensions | 128→128→1 | 128→128→1 |
| [Training Configuration] | | |
| Initial Learning Rate | $5\times10^{-4}$ ($1\times10^{-3}$)* | $5\times10^{-4}$ |
| Optimizer | Adam ($\beta_1$=0.9, $\beta_2$=0.999) | Adam ($\beta_1$=0.9, $\beta_2$=0.999) |
| Learning Rate Scheduler | Cosine Annealing | Cosine Annealing |
| Gradient Clipping | 1.0 | 1.0 |
| Batch Size | 128 | 128 |
| Epochs | 1,000 | 500 |
| Target Normalization | MAD Normalization | None |

*$1\times10^{-3}$ for orbital energy properties ($\varepsilon_{HOMO}$, $\varepsilon_{LUMO}$, $\Delta\varepsilon$); $5\times10^{-4}$ for other properties.

The QM9 dataset comprises molecular geometries optimized using the B3LYP/6-31G(2df,p) level of density functional theory (DFT) and includes 12 quantum chemical properties. Consistent with the original EGNN investigation, the dataset underwent random partitioning into a training set of 110,000 molecules, a validation set of 10,000 molecules, and the rest as a test set, utilizing a constant random seed. PFP descriptors were calculated utilizing the ωB97XD mode (v8). A fundamental four-dimensional vector is utilized for edge features.

In the EGNN layers, sum aggregation was utilized for message aggregation. Sum aggregation maintains extensivity, aligning with the characteristics of energy-related quantities like internal energy, enthalpy, and free energy, which scale linearly with the number of atoms. Because of the high optimization of the QM9 molecular structures at the B3LYP/6-31G(2df,p) level, additional structural refinement was considered unnecessary, and coordinate updates were deactivated during training.

In the output layer, sum pooling was applied to all 12 target properties. This design aligns with the original EGNN formulation and is especially suitable for energy-related properties, typically represented as the sum of atomic contributions. Sum pooling was similarly employed for the dipole moment μ, which is defined as the sum of products of atomic partial charges and positions.

The tmQM dataset comprises molecular properties calculated using single-point DFT at the TPSSh-D3BJ/def2-SVP level after semi-empirical structure optimization with GFN2-xTB. The dataset underwent stratified sampling, resulting in a training set of 70,000 complexes, a validation set of 8,000 complexes, and a test set of 8,000 complexes. PFP descriptors were computed using the r2SCAN mode (v8). For edge features, a five-dimensional vector was utilized and supplemented with qualitative bond-order information. Specifically, the Wiberg bond order ($BO_{ij} \in \mathbb{R}$) was derived from the tmQM dataset, computed at the GFN2-xTB level of theory. This descriptor quantitatively represents the strength of covalent bonds as a continuous value. For each edge $(i, j)$, the BO value was taken from the tmQM dataset when available; otherwise, $BO_{ij}$ was set to 0.0.

Mean aggregation was utilized for message aggregation in the EGNN layers of tmQM. This decision was driven by the varied coordination numbers observed in transition metal complexes within the tmQM dataset. The use of sum aggregation could introduce a bias dependent on the coordination number. Normalizing the aggregated messages by the number of neighbors ensured that messages were collected consistently across different coordination geometries, promoting more stable learning. Furthermore, as the tmQM structures are optimized using GFN2-xTB, which may be less accurate than fully DFT-optimized structures, coordinate updates are permitted during training to aid in structural refinement.

In the output layer for tmQM, sum pooling was applied to the dipole moment $\mu$, while mean pooling was used for orbital energies ($\varepsilon_{HOMO}$, $\varepsilon_{LUMO}$, $\Delta\varepsilon$) and the metal partial charge Metal_q. Mean pooling allows the electronic state of the entire molecule to be represented consistently on a scale independent of coordination number, facilitating the aggregation of electronic interactions between the metal center and ligands with equal weights. In particular, for the metal partial charge, the presence of exactly one metal atom in each complex was leveraged to predict the metal oxidation state, treating the entire coordination environment as a molecular-level aggregate.

The output layer architecture was consistent across the datasets. The node decoder $\phi_{node}$ comprises two layers with an input dimension of 128, a hidden dimension of 128, and an output dimension of 128. Similarly, the graph decoder has two layers with an input dimension of 128, a hidden dimension of 128, and an output dimension of 1. In both MLPs, the Swish activation function was applied after linear transformations in the hidden layers. However, the final layer did not use any activation function, leading to a linear transformation.

3.2 Training Configuration

A unified optimization strategy was adopted for both datasets. The primary hyperparameters are detailed in Table 2. The Adam optimizer was utilized with parameters $\beta_1 = 0.9$, $\beta_2 = 0.999$, $\varepsilon = 1\times10^{-8}$, and weight decay of $1\times10^{-16}$.[28] The initial learning rate was established at $1\times10^{-3}$ for orbital-energy-related properties in QM9 ($\varepsilon_{HOMO}$, $\varepsilon_{LUMO}$, $\Delta\varepsilon$), and at $5\times10^{-4}$ for all other properties as well as for all properties in tmQM. A higher learning rate was applied to orbital energies due to their broader numerical range, resulting in relatively smaller gradients in the loss function.

As a learning rate scheduler, cosine annealing is used to gradually decrease the learning rate in the later training stages, facilitating convergence to the local optima. The learning rate $\eta(t)$ is expressed as:

$$\eta(t) = \eta_{min} + \frac{1}{2}(\eta_{max} - \eta_{min})\left(1 + \cos\left(\frac{\pi t}{T}\right)\right), \quad (15)$$

where $\eta_{max}$ denotes the initial learning rate, $\eta_{min}$ the minimum learning rate, $t$ the current epoch, and T the total number of epochs. In this study, $\eta_{min}$ was set to $1\times10^{-7}$ and $T$ was set to 1,000. To prevent gradient explosion, gradient clipping with a maximum norm of 1.0 was applied to all parameters. The mean squared error was used as the loss function.

For the QM9 dataset, the target properties were standardized to facilitate consistent learning across properties with varying numerical scales. In contrast, target normalization was omitted for tmQM as the property values in this dataset exhibited relatively uniform scales.

Training was conducted using a batch size of 128 for a maximum of 1,000 epochs, implementing early stopping according to the validation loss. Independent models were trained for each target property. The computations were executed on an NVIDIA Tesla V100-PCIE-16GB GPU. The software stack included PyTorch 2.x, PyTorch Geometric 2.x, and Python 3.10. Random seeds were set at 42 to guarantee reproducibility.

The model performance was assessed using the mean absolute error (MAE), root mean squared error, and coefficient of determination ($R^2$).

3.3 Comparative Method

To evaluate molecular property prediction accuracy of the QM9 and tmQM datasets, we compared the suggested model with existing methods that vary in their treatment of 3D molecular configurations and the incorporation of equivariance. These approaches are categorized according to the nature of the geometric data utilized as input. Their key attributes are briefly summarized below.

A method based solely on distance information, SchNet[3] is a pioneering 3D molecular learning model that utilizes continuous filter convolutions to represent local atomic environments based on interatomic distances. As a method that explicitly incorporates angular information, DimeNet$^{++}$[29] introduces bond angles (three-body interactions) alongside interatomic distances to precisely represent molecular geometric constraints through directional message passing. SphereNet[11] efficiently processes directional information using spherical harmonics and encodes angular dependence in the embedding space.

As a method based on equivariant message passing, PaiNN5 utilizes equivariant layers to concurrently update scalar and vector features, leading to enhanced accuracy in predicting vector quantities such as dipole moments. ViSNet[13] improves the quality of geometric representations by reinforcing interactions between vectors and scalars within equivariant message passing.

As methods that handle higher-order equivariant representations, TFN[6] systematically treats scalars, vectors, and higher-order tensors through equivariant layers based on spherical harmonics. SE(3)-Transformer[7] extends the representational power by introducing equivariant attention mechanisms, while Cormorant[30] realizes SO(3)-equivariant higher-order tensor representations using Clebsch–Gordan products. L1Net[31] combines rotation-equivariant convolutions with local density representations, and LieConv[32] implements general equivariance through the convolutions of Lie groups.

As a strategy for thorough information integration, ComENet[12] examines interactions among all atom pairs via a full message-passing mechanism, while PaxNet[10] hierarchically handles local and nonlocal interactions using multiplex graphs.

For comparison with the tmQM dataset, we utilized two approaches that leveraged quantum chemical descriptors. NatQG[33] utilizes graph representations based on natural bond orbital (NBO) analysis and explicitly includes quantum-chemical information, such as orbital occupation numbers, hybridization states, and charge distributions as node features, enabling a comprehensive depiction of electronic states in transition-metal complexes. Additionally, QTAIM-GNN,[34] which is grounded in electron density topology, integrates information from quantum theory of atoms in molecules (QTAIM) analysis, encompassing atomic charges, bond critical points, and electron density gradients, into graphical structures. Various versions of QTAIM-GNN have been documented based on different levels of structural optimization and density calculation. In this investigation, we focus on the variant utilizing xTB-optimized structures and TPSS-D3BJ/def2-SVP electron densities. It is important to note that NatQG (tmQMg: approximately 60,000 molecules, PBE0-D3BJ/def2-TZVP labels) and QTAIM-GNN (tmQM+: approximately 60,000 molecules, xTB-optimized structures, TPSS-D3BJ/def2-SVP densities, PBE0-D3BJ/def2-TZVP labels) employ more advanced theoretical frameworks and distinct dataset sizes compared to the original tmQM dataset employed in this study (86,000 molecules, GFN2-xTB-optimized structures, TPSSh-D3BJ/def2-SVP labels). Hence, caution is advised when directly comparing the reported prediction accuracies.

The base model utilized in this study, the EGNN, is a streamlined architecture that achieves E(n) equivariance without relying on higher-order tensor representations, taking solely atomic numbers and coordinates as input. An EGNN baseline model, excluding PFP descriptors, was developed to evaluate the efficacy of these descriptors. In the QM9 baseline, node features were limited to one-hot encodings of atomic numbers (five dimensions), while edge features were restricted to interatomic distances (one dimension). Conversely, in the tmQM baseline, node features encompassed five-dimensional geometric attributes (atomic number, distance to the molecular center of mass, distance to the nearest atom, mean interatomic distance, and coordination number), with edge features comprising solely interatomic distances. Across all baseline models, the EGNN architecture and training parameters mirrored those of the proposed model, ensuring equitable comparisons.

4. Results and Discussion

In this section, we evaluate the predictive performance of the EGNN-PFP model on the QM9 and tmQM datasets and analyze its effectiveness by comparing it with current methods.

4.1 Prediction Results on the QM9 Dataset

We evaluated the molecular property prediction performance of the EGNN-PFP model using the QM9 dataset. We evaluated its efficacy by contrasting it with existing 3D GNN approaches and a baseline model lacking PFP descriptors. The energy-related properties in QM9 ($U_0$, $U$, $H$, $G$, and ZPVE) are comprehensive and exhibit systematic variations with molecular size. As a result, $R^2$ exceeds 0.99 for most models, diminishing its discriminatory power for performance evaluation. Therefore, this study focused on MAE to facilitate quantitative comparison of performance differences between models.

Table 3 summarizes the MAE values of the EGNN-PFP model and existing 3D GNN methods for all 12 properties in the QM9 dataset. EGNN-PFP achieved a prediction accuracy comparable to or better than that of the original EGNN for 11 of the 12 properties. Particularly, consistent performance improvements were observed for properties strongly dependent on electronic states, such as energy-related properties, frontier orbital energies, and dipole moments.

A comparison with the EGNN baseline without PFP descriptors clearly shows that PFP enhances the accuracy across various properties. For instance, the MAE of the dipole moment decreases by about 24%, from 0.029 D to 0.022 D. Similarly, the MAE for the HOMO energy decreases by around 21%, from 28.2 meV to 22.4 meV, and for the HOMO–LUMO gap, the error is reduced by about 15%, from 47.6 meV to 40.4 meV. These findings suggest that the pretrained PFP descriptors effectively capture local electronic state information and can be effectively utilized for predicting molecular properties in the QM9 dataset.

When analyzed by property type, the EGNN-PFP outperformed the original EGNN for all extensive energy-related properties ($U_0$, $U$, $H$, $G$, and ZPVE). For instance, the MAE values achieved by EGNN-PFP are 9.7 meV for $U_0$ (11 meV in the original EGNN), 10.1 meV for $U$ (12 meV), 10.5 meV for $H$ (12 meV), 10.5 meV for $G$ (12 meV), and 1.33 meV for ZPVE (1.55 meV), demonstrating consistent enhancements. As these properties consist of atomic contributions, the utilization of sum pooling is physically appropriate, and the outcomes indicate that the PFP descriptors accurately capture atomic-level energy contributions.

Table 3: MAEs for the prediction of 12 molecular properties on the QM9 dataset.

| Methods | $\mu$ (D) | $\Delta\varepsilon$ (eV) | $\varepsilon_{HOMO}$ (eV) | $\varepsilon_{LUMO}$ (eV) | $U_0$ (meV) | $U$ (meV) | $G$ (meV) | $H$ (meV) | ZPVE (meV) | $\alpha$ (Bohr$^3$) | $C_v$ (cal/mol·K) | $<R^2>$ (Bohr$^2$) |
|---|---|---|---|---|---|---|---|---|---|---|---|---|
| SchNet[3] | 0.033 | 63 | 41 | 34 | 14 | 19 | 14 | 14 | 1.70 | 0.235 | 0.033 | 0.073 |
| DimeNet$^{++}$[29] | 0.030 | 33 | 25 | 20 | 6 | 6 | 8 | 7 | 1.21 | 0.044 | 0.023 | 0.331 |
| SphereNet[11] | 0.025 | 31 | 23 | 19 | 6 | 6 | 8 | 6 | 1.12 | 0.045 | 0.022 | 0.268 |
| PaiNN[5] | 0.012 | 46 | 28 | 20 | 6 | 6 | 7 | 6 | 1.28 | 0.045 | 0.024 | 0.066 |
| ViSNet[13] | 0.010 | 22 | 17 | 15 | 4 | 4 | 6 | 5 | 1.56 | 0.041 | 0.023 | 0.030 |
| TFN[6] | 0.064 | 58 | 40 | 38 | - | - | - | - | - | 0.223 | 0.101 | - |
| SE(3)-Tr.[7] | 0.051 | 53 | 35 | 33 | - | - | - | - | - | 0.142 | 0.054 | - |
| Cormorant[30] | 0.038 | 61 | 34 | 38 | 22 | 21 | 20 | 21 | 2.03 | 0.085 | 0.026 | 0.961 |
| L1Net[31] | 0.043 | 68 | 46 | 35 | 13 | 14 | 14 | 14 | 2.28 | 0.088 | 0.031 | 0.354 |
| LieConv[32] | 0.032 | 49 | 30 | 25 | 19 | 19 | 22 | 24 | 1.21 | 0.084 | 0.038 | 0.800 |
| ComENet[12] | 0.025 | 32 | 23 | 20 | 7 | 7 | 8 | 7 | 1.20 | 0.045 | 0.024 | 0.259 |
| PaxNet[10] | 0.011 | 31 | 23 | 20 | 6 | 6 | 7 | 6 | 1.17 | 0.045 | 0.023 | 0.093 |
| EGNN[8] | 0.029 | 48 | 29 | 25 | 11 | 12 | 12 | 12 | 1.55 | 0.071 | 0.031 | 0.106 |
| EGNN* | 0.029 | 47.6 | 28.2 | 23 | 13.6 | 12.9 | 10.8 | 11.3 | 1.58 | 0.061 | 0.033 | 0.105 |
| **EGNN-PFP** | **0.022** | **40.4** | **22.4** | **21.0** | **9.7** | **10.1** | **10.5** | **10.5** | **1.33** | **0.060** | **0.027** | **0.168** |

*Baseline model without PFP descriptors (this work).

For frontier orbital energies ($\varepsilon_{HOMO}$, $\varepsilon_{LUMO}$, $\Delta\varepsilon$), EGNN-PFP consistently outperformed the original EGNN. Specifically, EGNN-PFP achieved MAEs of 22.4 meV (29 meV) for the HOMO energy, 21.0 meV (25 meV) for the LUMO energy, and 40.4 meV (48 meV) for the HOMO–LUMO gap. These properties are challenging to learn due to their wide range of values. However, the higher initial learning rates used in this study, combined with the local electronic state representation provided by the PFP descriptors, are believed to have contributed to efficient learning and improved accuracy.

For dipole moment $\mu$, EGNN-PFP achieved an MAE of 0.022 D, surpassing the original EGNN (0.029 D) by approximately 24%. The enhancement in accuracy can be attributed to the PFP descriptors capturing local electron density distribution information, which plays a crucial role in determining the dipole moment based on atomic partial charges and spatial configuration. Similarly, for polarizability $\alpha$ (0.060 Bohr³) and heat capacity $C_v$ (0.027 cal/(mol·K)), EGNN-PFP demonstrated roughly 15% improvement (compared to 0.071 Bohr³) and 10% enhancement (compared to 0.031 cal/(mol·K)) in accuracy over the original EGNN, respectively.

In contrast, for the electronic spatial extent $<R^2>$, EGNN-PFP (0.168 Bohr²) performs less effectively than both the baseline (0.105 Bohr²) and the original EGNN (0.106 Bohr²). The electronic spatial extent indicates the overall spatial distribution of electron density across the molecule and cannot be adequately captured by merely aggregating local atomic environment information. Instead, mechanisms explicitly modeling long-range electronic correlations on the molecular scale are necessary. This outcome implies that while PFP descriptors are highly efficient for properties reliant on local electronic states, the current architecture is inadequate for properties characterized by long-range spatial delocalization.

The EGNN-PFP exhibited strong competitiveness in comparison to existing 3D GNN methods. Compared to SchNet, EGNN-PFP significantly outperformed across all properties, including the dipole moment (0.022 vs. 0.033 D), HOMO–LUMO gap (40.4 vs. 63 meV), and internal energy $U$ (10.1 vs. 19 meV). This enhancement can be attributed to the explicit utilization of pretrained electronic-state information in EGNN-PFP, while SchNet relies solely on atomic numbers and distances. Although DimeNet[++] demonstrated competitive or superior performance for certain properties, EGNN-PFP achieved comparable or superior accuracy for essential properties such as the dipole moment and orbital energies. In addition, EGNN-PFP offers enhanced computational efficiency due to its exclusion of explicit handling of three-body interactions, which is a practical advantage. In comparison to the other methods detailed in Table 3, EGNN-PFP did not attain the performance level of state-of-the-art techniques such as PaiNN, ViSNet,

PaxNet, SphereNet, and ComENet for certain properties. However, it exhibited superior performance across numerous properties compared to SchNet, DimeNet[++], TFN, SE(3)-Tr., Cormorant, L1Net, and LieConv.

4.2 Prediction Results on the tmQM Dataset

Table 4 summarizes the MAE values for the dipole moment, HOMO–LUMO gap, HOMO energy, LUMO energy, and metal partial charge obtained using EGNN, EGNN-PFP, NatQG, and QTAIM-GNN on the tmQM dataset, which comprises transition-metal complexes. The y–y plots for each property predicted by EGNN and EGNN-PFP are depicted in Figure 2.

**Table 4: MAEs for the prediction of five molecular properties on the tmQM dataset.**

| Methods | $\mu$ (D) | $\Delta\varepsilon$ (eV) | $\varepsilon_{HOMO}$ (eV) | $\varepsilon_{LUMO}$ (eV) | Metal_q |
|---|---|---|---|---|---|
| NatQG[33] | 0.545 | 0.132 | - | - | - |
| QTAIM-GNN[34] | - | 0.20 | 0.14 | 0.16 | - |
| EGNN* | 1.447 | 0.267 | 0.147 | 0.158 | 0.0562 |
| **EGNN-PFP** | **0.540** | **0.128** | **0.082** | **0.090** | **0.0196** |

*Baseline model without PFP descriptors (this work).

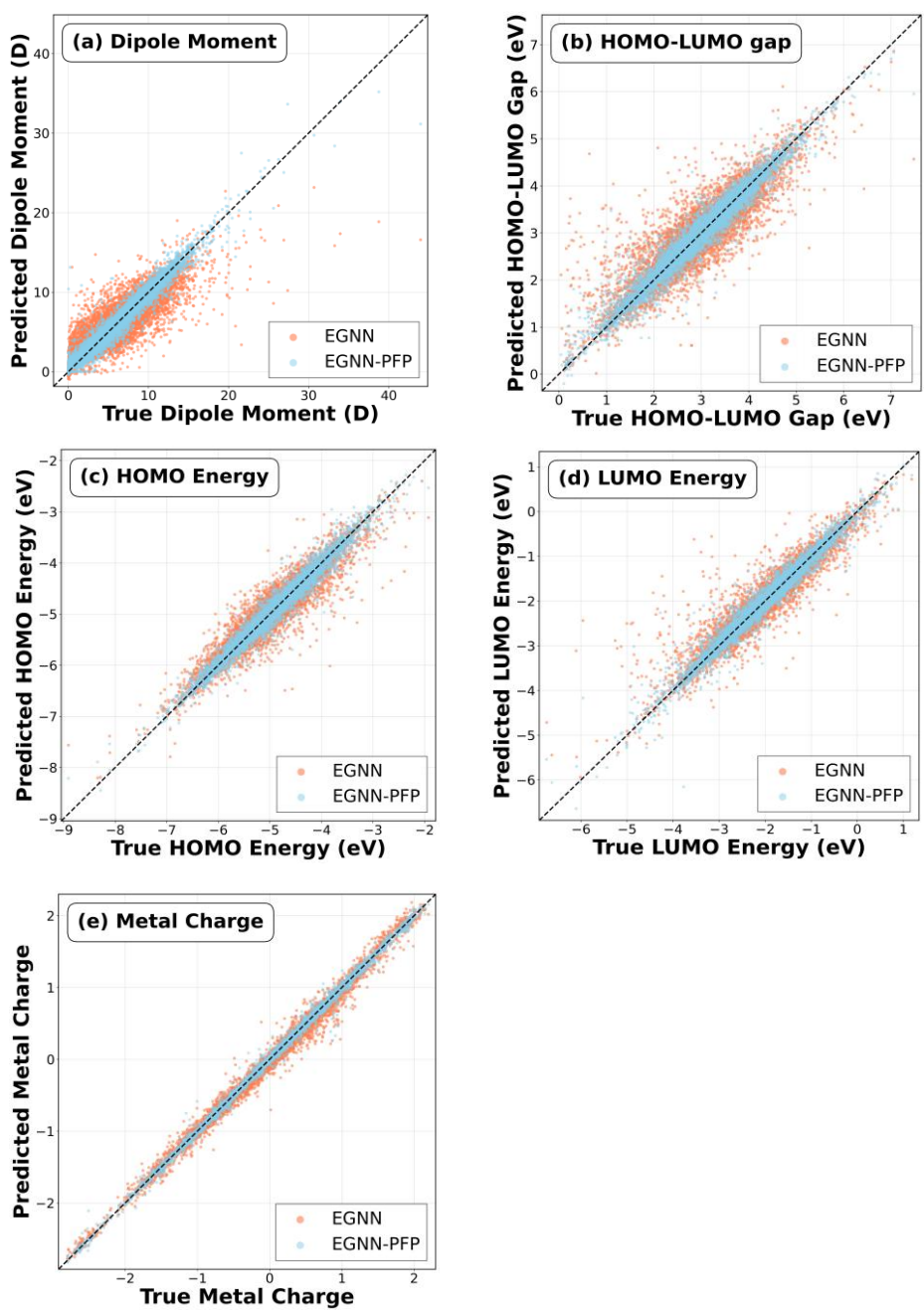

**Figure 2: Comparison of the baseline EGNN and EGNN-PFP models on the tmQM dataset for (a) dipole moment, (b) HOMO–LUMO gap, (c) HOMO energy, (d) LUMO energy, and (e) metal charge.**

In Figure 2, panels (a)–(e) correspond to the dipole moment, HOMO–LUMO gap, HOMO energy, LUMO energy, and partial metal charge, respectively. The horizontal axis represents the actual values, while the vertical axis represents the predicted values. The outcomes for the EGNN baseline are depicted in orange, while those for EGNN-PFP are illustrated in blue.

A comparison with the EGNN baseline without PFP descriptors clearly demonstrates that PFP improves the accuracy across all target properties on the tmQM dataset. Specifically, the MAE of the dipole moment was reduced by 63%, from 1.447 to 0.540 D. Errors reduced from 0.147 to 0.082 eV (44%), 0.158 to 0.090 eV (43%), 0.267 eV to 0.128 eV (52%), and 0.0562 to 0.0196 (65%) for the HOMO energy, the LUMO energy, HOMO–LUMO gap, and metal partial charge, respectively.

In the baseline model, the coefficients of determination were restricted to $R^2 = 0.716$ for the HOMO–LUMO gap and $R^2 = 0.790$ for the dipole moment. Conversely, the EGNN-PFP model, which includes PFP descriptors, showed significant enhancements, reaching $R^2 = 0.955$ and $R^2 = 0.950$ for these properties.

Transition metal complexes exhibit diverse coordination numbers and geometries, with intricate electronic states due to the involvement of d orbitals. As a result, the baseline model, relying solely on atomic numbers and geometric structures as input, struggles to fully represent the electronic environment near the metal centers. In contrast, incorporating PFP descriptors facilitates the precise depiction of potential fields surrounding metal centers and the transfer of charge between metals and ligands, resulting in significantly improved predictive capabilities.

The performance of EGNN-PFP is comparable to or better than that of existing methods, such as NatQG and QTAIM-GNN. However, these methods were trained on different datasets (approximately 60,000 molecules), whereas the dataset used in the present study contained approximately 86,000 molecules. Therefore, differences in the dataset size, composition, and data-splitting strategies should be considered when interpreting direct performance comparisons. Additionally, NatQG requires NBO analysis, and QTAIM-GNN requires QTAIM analysis for each molecule, both of which involve additional quantum chemical calculations. In contrast, EGNN-PFP can efficiently obtain PFP descriptors using the Matlantis API and does not require additional quantum chemical calculations.

The consistent performance improvements observed for both QM9 and tmQM datasets indicate that the PFP descriptors exhibit high flexibility with respect to elemental diversity. Many conventional molecular GNNs are optimized for specific element sets, primarily targeting organic molecules, and they typically require retraining for extension to new elements. By leveraging pretrained representations from Matlantis, which was developed as a general machine learning potential covering 96 elements, the EGNN-PFP can operate across a vast chemical space without additional tuning. The ability to achieve high-accuracy predictions using the same model architecture and training procedure, from QM9 with five elements (H, C, N, O, and F) to transition-metal complexes in tmQM containing more than 30 elements, represents a significant result demonstrating the effectiveness of transferring local descriptors derived from machine learning potentials pretrained on large-scale DFT calculations to molecular property prediction tasks.

5. Summary

In this study, we present a comprehensive framework for incorporating representations of local atomic environments acquired from machine learning potentials pretrained on large-scale quantum chemical data into tasks related to predicting molecular properties. While machine learning potentials inherently capture electronic and geometric details surrounding atoms as high-dimensional representations, the full potential of this information has not been maximized in GNNs for predicting molecular properties. Through the integration of pretrained local descriptors with GNNs, we methodically assessed the efficacy of this framework in enhancing the performance of molecular property prediction.

As an illustration, we developed an EGNN-PFP model by incorporating a 256-dimensional PFP pretrained with Matlantis into an EGNN. We assessed its performance on two significant quantum-chemical datasets comprising organic molecules and transition-metal complexes (QM9 and tmQM). On the QM9 dataset (133,831 molecules with 12 target properties), EGNN-PFP outperformed the original EGNN study for 11 out of 12 properties. In comparison to a baseline model lacking PFP descriptors, we observed error reductions of about 24%, 21%, and 15% for the dipole moment, HOMO energy, and HOMO–LUMO gap, respectively. This quantitatively illustrates that local descriptors derived from machine learning potentials effectively enhance molecular property prediction.

Furthermore, on the tmQM dataset (86,000 transition-metal complexes with five target properties), EGNN-PFP achieved significant improvements in accuracy across all properties. Specifically, error reductions of about 52% for the HOMO–LUMO gap (MAE $0.267 \rightarrow 0.128$ eV) and about 63% for the dipole moment ($1.447 \rightarrow 0.540$ D) were observed. These results illustrate that pretrained local descriptors remain effective even for transition metal complexes with intricate electronic structures involving d orbitals. The capability to attain high-accuracy predictions without additional tuning, from QM9, which includes five elements (H, C, N, O, and F), to tmQM, which encompasses complexes with over 30 elements, highlights the generality of the proposed approach.

Another significant implication of this study is that machine learning capabilities, traditionally focused on predicting energies and interatomic forces, can significantly broaden the scope of molecular properties that can be predicted with high accuracy by utilizing their internal representations. PFP descriptors are computed efficiently during the forward pass of machine learning models and do not require additional quantum chemical calculations. As a result, by reusing a single pretrained representation, a wide variety of molecular properties can be swiftly and precisely predicted within the computational framework of machine learning models.

The primary contribution of this study is not focused on enhancing a particular GNN architecture or improving the accuracy of a single property. Instead, it highlights the significance of transferring local electronic representations obtained from machine learning potentials to predict molecular properties, thereby expanding the output space. This approach, distinct from refining geometric representations, is anticipated to enhance the applicability of molecular property prediction when integrated with various evolving 3D GNN architectures.

Acknowledgements

References
(1) Gilmer, J.; Schoenholz, S. S.; Riley, P. F.; Vinyals, O.; Dahl, G. E. Neural Message Passing for Quantum Chemistry. Proc. Mach. Learn. Res. **2017**, *70*, 1263–1272.
(2) Butler, K. T.; Davies, D. W.; Cartwright, H.; Isayev, O.; Walsh, A. Machine Learning for Molecular and Materials Science. Nature **2018**, *559*, 547–555.
(3) Schütt, K. T.; Kindermans, P.-J.; Sauceda, H. E.; Chmiela, S.; Tkatchenko, A.; Müller, K.-R. SchNet: A Continuous-Filter Convolutional Neural Network for Modeling Quantum Interactions. Adv. Neural Inf. Process. Syst. **2017**, *30*, 992–1002.
(4) Gasteiger, J.; Groß, J.; Günnemann, S. Directional Message Passing for Molecular Graphs. In Proc. ICLR **2020**.
(5) Schütt, K. T.; Unke, O. T.; Gastegger, M. Equivariant Message Passing for the Prediction of Tensorial Properties and Molecular Spectra. Proc. Mach. Learn. Res. **2021**, *139*, 9377–9388.
(6) Thomas, N.; Smidt, T.; Kearnes, S.; Yang, L.; Li, L.; Kohlhoff, K.; Riley, P. Tensor

Field Networks: Rotation- and Translation-Equivariant Neural Networks for 3D Point Clouds. arXiv preprint arXiv:1802.08219, **2018**.

(7) Fuchs, F. B.; Worrall, D. E.; Fischer, V.; Welling, M. SE(3)-Transformers: 3D Roto-Translation Equivariant Attention Networks. Adv. Neural Inf. Process. Syst. **2020**, *33*, 1970–1981.

(8) Satorras, V. G.; Hoogeboom, E.; Welling, M. E(n)-Equivariant Graph Neural Networks. Proc. Mach. Learn. Res. **2021**, *139*, 9323–9332.

(9) Gasteiger, J.; Becker, F.; Günnemann, S. GemNet: Universal Directional Graph Neural Networks for Molecules. Adv. Neural Inf. Process. Syst. **2021**, *34*, 6790–6802.

(10) Zhang, S.; Liu, Y.; Xie, L. Physics-Aware Graph Neural Network for Accurate RNA 3D Structure Prediction. Machine Learning for Structural Biology Workshop, NeurIPS **2022**.

(11) Liu, Y.; Wang, L.; Liu, M.; Lin, Y.; Zhang, X.; Oztekin, B.; Ji, S. Spherical Message Passing for 3D Molecular Graphs. In Proc. ICLR **2022**.

(12) Wang, L.; Liu, Y.; Lin, Y.; Liu, H.; Ji, S. ComENet: Towards Complete and Efficient Message Passing for 3D Molecular Graphs. Adv. Neural Inf. Process. Syst. **2022**, *35*, 650–664.

(13) Wang, Y.; Li, S.; He, X.; Li, M.; Wang, Z.; Zheng, N.; Shao, B.; Liu, T.-Y.; Wang, T. Enhancing Geometric Representations for Molecules with Equivariant Vector–Scalar Interactive Message Passing. Nat. Commun. **2024**, *15*, 313.

(14) Behler, J. Atom-Centered Symmetry Functions for Constructing High-Dimensional Neural Network Potentials. J. Chem. Phys. **2011**, *134*, 074106.

(15) Bartók, A. P.; Kondor, R.; Csányi, G. On Representing Chemical Environments. Phys. Rev. B **2013**, *87*, 184115.

(16) Huo, H.; Rupp, M. Unified representation of molecules and crystals for machine learning. *Mach. Learn.: Sci. Technol.* **2022**, *3,* 045017.

(17) Drautz, R. Atomic Cluster Expansion for Accurate and Transferable Interatomic Potentials. Phys. Rev. B **2019**, *99*, 014104.

(18) Christensen, A. S.; Bratholm, L. A.; Faber, F. A.; von Lilienfeld, O. A. FCHL Revisited: Faster and More Accurate Quantum Machine Learning. J. Chem. Phys. **2020**, *152*, 044107.

(19) Smith, J. S.; Isayev, O.; Roitberg, A. E. ANI-1: An Extensible Neural Network Potential with DFT Accuracy at Force Field Computational Cost. Chem. Sci. **2017**, *8*, 3192–3203.

(20) Zhang, L.; Han, J.; Wang, H.; Car, R.; E, W. Deep Potential Molecular Dynamics: A Scalable Model with the Accuracy of Quantum Mechanics. Phys. Rev. Lett. **2018**, *120*,

143001.

(21) Unke, O. T.; Meuwly, M. PhysNet: A Neural Network for Predicting Energies, Forces, Dipole Moments, and Partial Charges. J. Chem. Theory Comput. **2019**, *15*, 3678–3693.

(22) Takamoto, S.; Shinagawa, C.; Motoki, D.; et al. Towards Universal Neural Network Potential for Material Discovery Applicable to an Arbitrary Combination of 45 Elements. Nat. Commun. **2022**, *13*, 2991.

(23) Batzner, S.; Musaelian, A.; Sun, L.; et al. E(3)-Equivariant Graph Neural Networks for Data-Efficient and Accurate Interatomic Potentials. Nat. Commun. **2022**, *13*, 2453.

(24) Batatia, I.; Kovacs, D. P.; Simm, G.; Ortner, C.; Csányi, G. MACE: Higher Order Equivariant Message Passing Neural Networks for Fast and Accurate Force Fields. Adv. Neural Inf. Process. Syst. **2022**, *35*, 11423–11436.

(25) Musaelian, A.; Batzner, S.; Johansson, A.; Sun, L.; Owen, C. J.; Kornbluth, M.; Kozinsky, B. Learning Local Equivariant Representations for Large-Scale Atomistic Dynamics. Nat. Commun. **2023**, *14*, 579.

(26) Ramakrishnan, R.; Dral, P. O.; Rupp, M.; von Lilienfeld, O. A. Quantum Chemistry Structures and Properties of 134k Molecules. Sci. Data **2014**, *1*, 140022.

(27) Balcells, D.; Skjelstad, B. B. tmQM Dataset—Quantum Geometries and Properties of 86k Transition Metal Complexes. J. Chem. Inf. Model. **2020**, *60*, 6135–6146.

(28) Kingma, D. P.; Ba, J. Adam: A Method for Stochastic Optimization. In Proc. ICLR **2015**.

(29) Klicpera, J.; Giri, S.; Margraf, J. T.; Günnemann, S. Fast and Uncertainty-Aware Directional Message Passing for Non-Equilibrium Molecules. NeurIPS ML4Molecules Workshop **2020**.

(30) Anderson, B.; Hy, T.-S.; Kondor, R. Cormorant: Covariant Molecular Neural Networks. Adv. Neural Inf. Process. Syst. **2019**, *32*, 14537–14546.

(31) Miller, B. K.; Geiger, M.; Smidt, T. E.; Noé, F. Relevance of Rotationally Equivariant Convolutions for Predicting Molecular Properties. arXiv preprint arXiv:2008.08461, **2020**.

(32) Finzi, M.; Stanton, S.; Izmailov, P.; Wilson, A. G. Generalizing Convolutional Neural Networks for Equivariance to Lie Groups on Arbitrary Continuous Data. arXiv preprint arXiv:2002.12880, **2020**.

(33) Kneiding, H.; Lukin, R.; Lang, L.; Reine, S.; Pedersen, T. B.; De Bin, R.; Balcells, D. Deep Learning Metal Complex Properties with Natural Quantum Graphs. Digital Discovery **2023**, *2*, 618–633.

(34) Gee, W.; Doyle, A.; Vargas, S.; Alexandrova, A. N. Multi-Level QTAIM-Enriched Graph Neural Networks for Resolving Properties of Transition Metal Complexes.

Digital Discovery **2025**, *4*, 3378–3388.